Japan.
\documentstyle[a4,12pt]{article}
\def\av#1{\langle #1 \rangle}

\def\expect#1{{\rm E}\left\{ #1 \right\}}
\def\td{\tau_{\rm dep}}
\def\ti{\tau_{\rm int}}
\def\te{\tau_{\rm eff}}

\begin{document}

\rightline{\large OUCMT-94-2}
\vspace*{5mm}

\huge
\leftline{\bf Statistical Dependence}
\leftline{\bf and Related Topics\footnote{
\rm to be published in

\sl Computer Simulations in Condensed Matter Physics VII,\\
\rm ed. D.~P. Landau, K.~K. Mon and H.~B. Sch\"uttler (Springer)}
}
\vspace*{1cm}
\Large
\leftline{\sl Macoto KIKUCHI $^1$ Nobuyasu ITO$^2$ and Yutaka
OKABE$^3$}
\vspace*{8mm}
\large
\leftline{$^1$Department of Physics, Osaka University,
Toyonaka 560, Japan}
\leftline{ e-mail: kikuchi@godzilla.kek.jp}
\leftline{$^2$Department of Applied Physics,  The University of
Tokyo,}
\leftline{ Bunkyo-ku, Tokyo 113, Japan}
\leftline{ e-mail: ito@sunflower.t.u-tokyo.ac.jp}
\leftline{$^3$Department of  Physics,  Tokyo Metropolitan University,
}
\leftline{ Hachioji, Tokyo 192-03, Japan}
\leftline{ e-mail: okabe@phys.metro-u.ac.jp}

\vspace*{2cm}
\normalsize
\noindent
{\bf Abstract:}

On the basis of the dynamical interpretation of Monte Carlo
simulations,
we discuss the relation of the equilibrium relaxation time,
the susceptibility and the statistical error.
We introduce a new quantity called {\it the statistical dependence
time}
$\tau_{dep}$, which gives the reduction factor for the statistical
degree of
freedom due to the dynamical correlations between the data.
A new method is proposed for calculating equilibrium relaxation time
using $\tau_{dep}$,
the method which does not require knowledge of any time-displaced

correlation function.
We apply this method to the critical dynamics of Ising models in two
and
three dimensions,
and estimate the dynamical critical exponent $z$ precisely.
Systematic errors in response functions due to short simulations
are also discussed from the viewpoint of the statistical dependence.
\newpage
\section{Introduction}

Although Monte Carlo methods are stochastic methods,
sequences obtained by simulations are often
regarded as realistic dynamical sequences of a model in concern;
And several dynamical properties, both in equilibrium and in
non-equilibrium,
are studied using Monte Carlo simulations.
Dynamical correlations in equilibrium are characterized by
the equilibrium relaxation time $\tau$.
Any correlations, both spatial and temporal,
in a physical quantity can be investigated using

the appropriate correlation functions.
In fact, the relaxation time $\tau$ is usually estimated from

the asymptotic behavior

of the corresponding time-displaced correlation function.
On the other hand, existence of the characteristic time scale $\tau$
means that
the measurements made in a simulation are not statistically
independent
of each other.
The statistical dependence here is thus a different aspect of
the dynamical correlation.
According to above considerations,
we expect that alternative approaches for studying dynamics are
possible,
which may be called {\em methods of statistical dependence analysis}.

Recently, we have proposed a new method for calculating $\tau$,
based on the statistical dependence analysis.\cite{KI93}
This method no longer requires calculation of time-displaced
correlation
functions;
Rather, it uses only a ratio of equilibrium averages
of the susceptibility and the statistical error of
a thermodynamic quantity in concern.
As a result,
it enables us to make statistical analyses in an unambiguous manner;
We can simply follow the standard analysing methods used for
static quantities.
It is in contrast with the conventional method,
where unambiguous statistical analyses are not easy to be done,
since the statistical errors in the time-displaced

correlation function for different time are usually

not independent of each other.
In fact, if all the data in a time-displaced correlation function
were simply regarded as independent ones,
the resulting statistical error in $\tau$
would be systematically underestimated.

The aim of the present paper is to review recent progress in
the statistical dependence method for studying dynamics.
The topics we will deal with are the estimation of the dynamical
critical

exponents of the Ising models in two and three dimensions,
and the reexamination of unbiased estimators for response functions.
Before proceeding to these topics,

we will describe the statistical dependence method in the next
section.

\section{The Statistical Dependence Time}

Let us start from discussing the relation of the response function
({\it e.g.} magnetic susceptibility for magnetic models),
the relaxation time, and the statistical error.
Roughly and intuitively speaking, the square of the statistical error
for
a thermodynamic quantity is proportional to

the corresponding susceptibility divided by $\tau$.
That relation was explored
in their pioneering work by M\"uller-Krumbhaar and Binder\cite{MB73}
back in 1973, although in a different context.
As will be shown later, however, the expression they obtained was
valid only
in the long-time limit.
In this section, we will give an expression which is correct
in any range of time.
The new method for calculating $\tau$ will be proposed in the course
of the
discussion.

Suppose we make $N$ independent simulations of a system
under the same parameters ({\it e.g.} temperature),
with measurements of a physical quantity $Q$ made in constant
time intervals;

This interval is used as a unit of time in the following.
Number of such measurements in each run is $n$.
Under this situation, we can define two different kinds of (thermal)
averages
as follows:
\begin{enumerate}
\item The average over the measurements in a single run,
\begin{equation}
 \av Q _\alpha \equiv  {1 \over n}\sum_{k=1}^{n} Q_\alpha(k),
\end{equation}
\item The average over the independent runs,
\begin{equation}
 \overline{\av Q} \equiv {1 \over N}\sum_{\alpha =1}^{N} \av Q
_\alpha,
\end{equation}
\end{enumerate}
where $Q_\alpha(k)$ denotes the value of the quantity $Q$ at the
$k$-th
measurement in the $\alpha$-th run, where $k=1$, $2$, $\cdots$, $n$
and
$\alpha =1$, $2$, $\cdots$, $N$.
The situation is schematically illustrated in Fig.~\ref{fig:setup}.

\begin{figure}[htb]
\rule{\textwidth}{0.1pt}
\caption{Schematic illustration of the simulation setup.

Each simulation runs horizontally.
The circles represent the measurements of a quantity $Q$;
$n$ measurements are made in each run.

The average $\av Q _{\alpha}$ is calculated from
all the measurements in $\alpha$ -th run.
The above procedure is repeated $N$ times independently.}
\label{fig:setup}
\rule{\textwidth}{0.1pt}
\end{figure}

It is well known in elementary statistics that

if all the $n$ data are statistically independent,
the statistical degrees of freedom (DOF) is $n$,
and the expectation value for the variance of $\av Q$

(we omitted the suffix $\alpha$ here,

since the expectation value does not depend on it) is

\begin{equation}
\expect{(\av Q - \mu)^2} = \frac{\sigma^2}{n}, \label{eq:indep}
\end{equation}
where $\mu \equiv \expect Q$ and
$\sigma^2 \equiv \expect{Q^2} - \mu^2$ are the expectation values of
$Q$
and of the variance of $Q$, respectively.
On the other hand, if the data are correlated to each other,
as is usually the case in dynamical simulations,
rather than they are independent,
the DOF will be reduced due to the mutual correlations.
Such reduction in the DOF can be discussed in terms of the
time-displaced correlation function of $Q$.
In fact, the expectation value for the variance of $\av Q$ for
the correlated data is given as

\begin{equation}
\expect{(\av Q - \mu)^2} =

 {1 \over {n^2}} \sum_{k, l = 1}^{n} \left( \expect{m(k) m(l)} -
\mu^2\right)
 = {1 \over {n^2}} \sum_{k, l = 1}^{n} \expect{C(k,l)},
\end{equation}
where $C(k,l)$ is the unnormalized time-displaced correlation
function of
the fluctuation in $Q$ for ``time'' $k$ and $l$.
Let us assume
that the relaxation of this correlation function is described

by a single exponential function, as is often the case,
with the associated relaxation time $\tau$;
That is, we assume that the correlation function is expressed as

\begin{equation}

\expect{C(k,l)} = \sigma^2 e^{-|k-l|/\tau}.
\end{equation}
Then we get

\begin{equation}
  \expect{(\av Q - \mu)^2}

=  {\sigma^2 \over n^2} \sum_{k, l=1}^{n} e^{-|k-l|/\tau}.
\label{eq:sum}
\end{equation}
If we approximate the summation in this equation by an integration
and
take the integration range as $[0,\infty)$,
we have

\begin{equation}
  \expect{(\av Q - \mu)^2} = \frac{2\tau}{n}\sigma^2.
\label{eq:intapp}
\end{equation}
The order of the thermal average and the long-time limit can not be
interchanged here,
because $\lim_{n\rightarrow\infty}\av Q = \mu$ by definition.
Equation~(\ref{eq:intapp}) is just the expression obtained

by M\"uller-Krumbhaar and Binder.
It should be noted that the limit $1/\tau\rightarrow 0$ is implicitly
assumed in the integral approximation as well as the explicit
long-time limit;
Equation~(\ref{eq:intapp}) is thus valid only in this limit.

Let us proceed further, without assuming the long-time limit.
The summation in Eq.~(\ref{eq:sum}) can be calculated explicitly, and
gives

\begin{equation}
\sum_{k, l=1}^{n} e^{-|k-l|/\tau}

= n  \left[ {{1+\Lambda}\over{1-\Lambda}}

      - {{2 \Lambda(1-\Lambda ^n)}\over{n(1-\Lambda)^2}}\right],
\end{equation}
where $\Lambda \equiv \exp (-1/\tau)$.
Then Eq.~(\ref{eq:sum}) can be expressed in a similar form as
Eq.~(\ref{eq:indep}) and (\ref{eq:intapp}) as follows:

\begin{equation}
 \expect{(\av Q - \mu)^2} = {{2 \td}\over n} \sigma^2,
\label{eq:reduc}
\end{equation}
with
\begin{equation}
\td = {1\over 2}\left[{{1+\Lambda}\over{1-\Lambda}}
      - {{2\Lambda (1-\Lambda ^n)}\over{n(1-\Lambda)^2}}\right].
\label{eq:td}
\end{equation}
We introduced a quantity $\td$ having the dimension of time;

We call it {\it the statistical dependence time}.
Comparing Eq.~(\ref{eq:reduc}) with Eq.~(\ref{eq:indep}), we see that
the factor $n/2\td$ plays a role of the DOF for correlated data;
The DOF is thus reduced by a factor $1/2\td$
due to the dynamical correlations.
In other words, only $n/2\td$ among $n$ measurements have
statistical significance,
and $2\td$ is interpreted as the mean interval of
successive statistically significant measurements.
It should be noted that this reduction factor $1/2\td$ depends not
only

on $\tau$ but also on $n$,
while in the conventional argument based on Eq.~(\ref{eq:intapp})
the reduction factor is just $1/2\tau$ irrespective of $n$.

The statistical dependence time $\td$ can be estimated by simulations
using Eq.~(\ref{eq:reduc}) as follows:
The unbiased estimator $(\delta Q)^2$ for $\expect{(\av Q - \mu)^2}$

is calculated from $N$ independent runs as

\begin{equation}
(\delta Q)^2

= {N\over{N-1}}\left( \overline{\av Q^2} - \overline{\av Q}^2
\right).
\label{eq:errest}
\end{equation}
 The quantity $|\delta Q|/\sqrt{N}$
is commonly used as the estimator for the statistical error in $Q$.

The variance $\sigma^2$, on the other hand, is calculated as
the fluctuation $\chi_Q$, which is
physically proportional to the response function associated to $Q$:

\begin{equation}
\chi_Q = \overline{\av{Q^2}} - \overline{\av Q}^2, \label{eq:susest}
\end{equation}
Combining these equations, we get our final expression for the
estimator of $\td$:

\begin{equation}
\td = {{n (\delta Q)^2}\over{2 \chi_Q}}
   = {{nN}\over {2(N-1)}}
     {{\overline{\av Q^2} - \overline{\av Q}^2 }\over
      {\overline{\av{Q^2}} - \overline{\av Q}^2 }}. \label{eq:tdest}
\end{equation}
Therefore, what is proportional to the square of the statistical
error
is not $\tau$ but $\td$.
Once $\td$ is thus estimated, we can then get $\tau$ by solving
Eq.~(\ref{eq:td}).

The above discussions can be generalized to cases of
multi-exponential relaxation,

\begin{equation}

\expect{C(k,l)} = \sum_I a_I e^{-|k-l|/\tau_I}, \label{eq:multi}
\end{equation}
where $\tau_I$ is the relaxation time of the $I$-th mode,
and $a_I$ the corresponding amplitude.
Then the statistical dependence time becomes
\begin{equation}
\td = \sum_I {{a_I}\over 2}\left[{{1+\Lambda_I}\over{1-\Lambda_I}}
      - {{2\Lambda_I (1-\Lambda_I^n)}
      \over{n(1-\Lambda_I)^2}}\right], \label{eq:tdmulti}
\end{equation}
where $\lambda_I = \exp (-1/\tau_I)$.
For practical use, however, it is convenient to follow the same
procedure

as in the case of single-exponential relaxation;
That is, we calculate the {\em effective} relaxation time $\te$

by solving Eq.~(\ref{eq:td}) instead of Eq.~(\ref{eq:tdmulti}).
The effective relaxation time $\te$ thus estimated

can, in fact, be used as a substitute for the
so-called integrated relaxation time $\ti$,
whenever the integrated relaxation time has a definite meaning.
Thus it obeys the same scaling law as $\ti$ at criticality.
More detailed discussions are given in Ref.~\cite{KI93}.

\section{Dynamical Critical Exponent of 3D Ising Model}

In this section, we present the results of applying the method
described in
the previous section to the critical dynamics of the
three-dimensional
Ising model.  The Hamiltonian is given as follows:

\begin{equation}
H = -J\sum_{\langle i,j\rangle} \sigma_i\sigma_j, \quad ( \sigma =
\pm 1)
\end{equation}
where the interactions act on the nearest-neighbor spin pairs.
At the critical point, the relaxation time obeys the finite-size
scaling law
with the dynamical critical exponent $z$:

\begin{equation}
\tau \sim L^z, \label{eq:dscale}
\end{equation}
where $L$ is the linear dimension of the system.
Therefore, by calculating $\tau$ for several system sizes at the
critical
point,  we can estimate $z$.

We performed Monte Carlo simulations for the Ising model on

the simple-cubic lattices of the size $L\times L\times (L+1)$,
with the skewed boundary conditions imposed in the $L$--directions,
and the periodic boundary condition in the $(L+1)$--direction.
We employed the Metropolis transition probability.
The Simulations were fully vectorized by means of
the two-interpenetrating-sublattice updating procedure.
A multispin coding technique were also used,

which assign a spin to a single bit.
The Simulations were made at the critical point $J/T_c = 0.221654$,
with 64 independent runs made simultaneously
by means of the recycling algorithm.\cite{IKO93}
We took 19 different system sizes, among which the smallest one was
$L=3$, and
the largest one was $L=91$.

We measured the magnetization and its square by the simulations,

and calculated the associated $\te$ from $\td$
by solving Eq.~(\ref{eq:td}).
Length of the simulations were adjusted to be about $10\td$,
with the same length discarded for equilibration.
The number of the independent runs $N$ was taken
from $23\times 64$ at smallest to $1400\times 64$ at largest,
depending on the system size.
These procedures were repeated several times for estimating the
statistical
errors in $\te$.
Figure~\ref{fig:3d} shows the log-log plot of $\te$ against
the {\em average} linear dimension $L_{\rm av} \equiv
\left[L^2(L+1)\right]^{1/3}$.
\begin{figure}[htb]
\rule{\textwidth}{0.1pt}
\caption{Log-log plot of $\te$ against
$L_{\rm av} \equiv\left[L^2(L+1)\right]^{1/3}$ for 3D Ising model.
The errors are much smaller than the symbols.
The straight line represents $z=2.030$.}
\label{fig:3d}
\rule{\textwidth}{0.1pt}
\end{figure}
As can be seen in the figure, the data fit very well to a straight
line.
But we found by detailed statistical analysis that $\te$ for $L=3$
deviated significantly from the straight line.
This deviation can be seen clearly in Fig.~\ref{fig:3dfit8}
where we show $z$ calculated from eight successive data;
\begin{figure}[htb]
\rule{\textwidth}{0.1pt}
\caption{The value of $z$ calculated from the data sets,

each of which consists of eight successive sizes.
The horizontal axis shows the mid value of $L_{\rm av}$ for each
set.}
\label{fig:3dfit8}
\rule{\textwidth}{0.1pt}
\end{figure}
That is, we took a set of eight successive data points in
Fig.~\ref{fig:3d},

and fitted them to a straight line;
The values of $z$ estimated that way for all the possible sets
are shown in the figure against
the mid value $L_{\rm mid}$ of $L_{\rm av}$ for each set.
The error bars indicate $1\sigma$ of the fit.
We see in the figure that only $z$ estimated from the set including
$L=3$ gives an extraordinarily large value,
while all the others fall in the range $2.030\pm 0.005$.
We confirmed also from the $\chi^2$ value of the fit

that only $\te$ for $L=3$ does not fit to the straight line.
Based on these statistical observations,

we can safely exclude $L=3$ from further analyses.
The least-square fit of the rest 18 data gave $z = 2.030\pm 0.004$;

Here, we took $2\sigma$ of the fit as the error
instead of $1\sigma$, just for the sake of safety.
The corresponding $\chi^2/{\rm DOF}$ of the fit was 1.05;
Thus the fitting of the data to the straight line is
extremely good.
Besides the statistical error, there may be the error due to
uncertainty
in $T_c$;\cite{I93}
Taking that error into account, our final estimate is
$z = 2.03\pm 0.01$, which is still quite accurate one.

Since the dynamics of three-dimensional models are equivalent to
the statics of some four-dimensional models,
which will be highly anisotropic in the fourth dimension,
it is sometimes expected that the dynamical scaling law for
three-dimensional
models has a logarithmic correction.
We, however, have not found any indication of the logarithmic
correction
to the dynamical finite-size scaling law, Eq.~(\ref{eq:dscale})
within the present accuracy;
Indeed, no particular tendency can be seen in
the residuals of the fit, which we plot in Fig.~\ref{fig:resid}
against

the system size.
\begin{figure}[htb]
\rule{\textwidth}{0.1pt}
\caption{The residuals of the fit for 3D Ising model against the
system size.
The values shown are not weighted by the statistical errors.}
\label{fig:resid}
\rule{\textwidth}{0.1pt}
\end{figure}

\section{Dynamical Critical Exponent of 2D Ising Model}

Dynamical critical exponent of two-dimensional Ising Model was
also calculated in the same line as in the previous section.
We made simulations of square-lattice Ising model
of the size $L(L+1)$ with the skewed boundary condition in
$L$--direction and the periodic boundary condition in another
direction.
Again we calculated the magnetization and its square by the
simulations
at the critical point $J/T_c = 0.4406868$, and estimated $\te$.
The same method of simulations as in the previous section was used.
The smallest system size was $L=11$, and
the largest one was $L=201$.
Again the length of the simulations were adjusted to be about
$10\td$,
with the same length discarded for equilibration.
The number of the independent runs $N$ was taken
from $64$ at smallest to $4000\times 64$ at largest,
depending on the system size.

Figure~\ref{fig:2d}
shows the log-log plot of $\te$ against
$L_{\rm av} \equiv \left[L(L+1)\right]^{1/2}$.
\begin{figure}[htb]
\rule{\textwidth}{0.1pt}
\caption{Log-log plot of $\te$ against
$L_{\rm av} \equiv\left[L(L+1)\right]^{1/2}$ for 2D Ising model.
The errors are much smaller than the symbols.
The straight line represents $z=2.173$.}
\label{fig:2d}
\rule{\textwidth}{0.1pt}
\end{figure}
Again, the data seem to fit very well to a straight line.
We follow similar statistical analyses made in the previous section.
Figure~\ref{fig:2dfit4} shows $z$ calculated from the data sets of

four successive sizes (not eight, this time) against $L_{\rm mid}$.
\begin{figure}[htb]
\rule{\textwidth}{0.1pt}
\caption{The value of $z$ calculated from the data sets,

each consisting of four successive sizes
against the mid value of $L_{\rm av}$ for each set.}
\label{fig:2dfit4}
\rule{\textwidth}{0.1pt}
\end{figure}
Unlike the three-dimensional case, the value of $z$ approaches slowly
to its
asymptotic value.
Even so, we can regard that the five data from $L=31$ to $L=201$
are in the asymptotic region as is seen in Fig.~\ref{fig:2dfit4}.
We confirmed that also from $\chi^2/{\rm DOF}$ of the fit.
By the least-square fit of these five data to the finite-size scaling
form
Eq.~\ref{eq:dscale},

we estimated $z=2.173\pm 0.016$, again with the error of $2\sigma$.
The value of $\chi^2/{\rm DOF}$ of the fit is 0.86;
Thus the data are fitted very well to a straight line.

\section{On unbiased estimators of response functions}
In this section,

we reexamine the arguments given by Ferrenberg, Landau, and
Binder\cite{FLB91}
on systematic errors in response function measurements,
from the viewpoint of the statistical dependence time.
Unlike the previous sections, we consider the case where only a
single run of a simulation is made.
It is well known that the unbiased estimator for the variance,
that is, the second-order cumulant of $Q$
from $n$ {\em independent} measurements is given as

\begin{equation}
\chi_Q = {n\over{n-1}}\left( \av Q^2 - \av Q^2 \right).
\label{eq:cumulant}
\end{equation}
The factor $n/(n-1)$ appears here for correction of the bias on the
estimate
due to the finite number of the measurements.
This correction factor has already been taken into account in
Eq.~(\ref{eq:errest}) in Sec.~1.
{}From the statistical-physics points of view,
the variance $\chi_Q$ is interpreted as the response function
associated to $Q$.
Therefore, the estimators for response functions
calculated from the fluctuations in general
should also be corrected by the above factor;
Otherwise, the measured response functions will be systematically
underestimated.
In actual simulations, however, the measurements are not independent

of each other, as has been discussed so far.
Therefore, we can not simply take $n/(n-1)$ as the correction factor;
The correction factor should be larger than it, due to the dynamical
correlations.

The above systematic errors due to finite-length
simulations have been pointed out in ref.~\cite{FLB91}
They argued that the response function estimated from

$n$ measurements $\chi_Q(n)$
and its true expectation value $\chi_Q(\infty)$ is related as

\begin{equation}
\chi_Q(n) = \chi_Q(\infty ) \left(1-\frac{2\tau}{n} \right).
\label{eq:tscale}
\end{equation}
Near the critical point, on the other hand,

several response functions exhibit scaling behaviors.
For example, the magnetic susceptibility $\chi_m$ at the critical
point
is scaled by the system size as

\begin{equation}
\chi_m(\infty) \sim L^{\gamma /\nu}, \label{eq:fsscalechi}
\end{equation}
where $\gamma$ and $\nu$ are the critical exponents of

the susceptibility and the correlation length, respectively.
Combining these two scaling relations, Eq.~(\ref{eq:tscale}) and

Eq.~(\ref{eq:fsscalechi}), they proposed a brand new scaling
relation,
which may be called the finite-size finite-measurement scaling,

for the susceptibility as

\begin{equation}
\chi_m(n) = a L^{\gamma /\nu} \left(1-\frac{2\tau}{n} \right).

\label{eq:newscale}
\end{equation}

However, as we have discussed in Sec. 2,
the statistical DOF for the correlated measurements are not $n/2\tau$
but $n/2\td$ in general.
Therefore, the scaling relation Eq.~(\ref{eq:newscale}) should be
corrected as follows:

\begin{equation}
\chi_m(n) = a L^{\gamma /\nu} \left(1-\frac{2\td}{n} \right).

\label{eq:newscale2}
\end{equation}
Equation~(\ref{eq:newscale2}) is expected be valid in general,
while Eq.~(\ref{eq:newscale}) is valid only when $n/2\tau$
is sufficiently large.
\begin{figure}[htb]
\rule{\textwidth}{0.1pt}
\caption{The finite-size finite-measurement scaling plot for
the magnetic susceptibility of the 3D Ising model at the critical
point.
The symbols are taken from the simulations by
Ferrenberg {\it et al.};{\protect \cite{FLB91}}
The data for several sizes from $L=16$ to $96$ are plotted.
The solid line represents the function $a (1-2\td/n)$,
while the dashed line represents the function $a (1-2\tau/n)$.
}
\label{fig:flbscale}
\rule{\textwidth}{0.1pt}
\end{figure}

In Fig.~\ref{fig:flbscale} we show the scaling plot of $\chi_m(n)$
for the three-dimensional Ising model at $J/T_c = 0.221654$;
The symbols represent the data
taken from the simulations in ref.~\cite{FLB91};
The two scaling functions $1-2\tau/n$ and $1-2\td/n$ are also shown
by the dashed line and the solid line, respectively,
with only the overall factor $a$ being the fitting parameter.
We see clearly that the scaling function $1-2\td/n$ agrees very well
to the simulation data.
On the other hand, the scaling function  $1-2\tau/n$ coincides with
the
data only for large $n/2\tau$, as is expected.
It should be noted that since the effective DOF $n/2\td$ depends not
only on
$\tau$ but also on $n$,
so does the scaling function $1-2\td/n$.
We, however, found that as long as $n$ is much larger than $10\tau$,
the scaling function does not depends on $n$ in the range of
$n/2\tau$
shown in the figure.

\section{Conclusions}
So far we have discussed the recent progress in the statistical
dependence
method for studying dynamics.
In the discussions in Sec.~2, we did not use any special property
inherent in Monte Carlo simulations;
Rather they are quite general, so that they
can apply also to other sorts of dynamical simulations,
such as the molecular dynamics method.

In Secs.~3 and 4, we have applied the statistical dependence time
method
for estimating the dynamical critical exponent $z$.
Although the relaxations at the critical point are in general
multi-exponential ones,
we found that the present method worked very well.
The dynamical critical exponents $z$ for the Ising model both in
two and three dimensions have been calculated by a lot of authors
so far.\cite{I93}
In case of three-dimensional Ising model,
the estimates using the equilibrium correlation time
spread within the range $1.99 \sim 2.17$.
Among them, recent two estimates based on large-scale

Monte Carlo simulations have given mutually incompatible values:
$2.04\pm 0.03$ by Wansleben and Landau\cite{WL91}
and $2.10 \pm 0.02$ by Heuer.\cite{H92}
Our estimation agrees well with the former and disagrees with the
latter
within the range of error.
Moreover, the statistical error in $z$ in the present result

is much smaller than that of the others.
It should be noted that we did not use any special technique

for estimating the statistical error in $z$;
We just followed the standard methods used for estimating the errors
in static quantities.
On the other hand, the error estimation must be more or less
complicated
and thus necessarily ambiguous in the conventional methods.
In case of the two-dimensional Ising model,
the estimates of $z$ so far are also scattering in a wide range.
The statistical error quoted for the present result
is one of the smallest errors among them.

In Sec.~4 we reexamined the argument on the systematic errors in
response functions due to short simulations by Ferrenberg {\it et
al.}.
We presented the correct scaling function for the scaling relation
they proposed.
Consequently, we confirmed both
the finite-size finite-measurement scaling they have proposed
and our proposal that $n/2\td$ gives the statistical DOF for
correlated data,
at the same time.
It should be noted the following:
In case of the magnetic susceptibility,
or response functions associated to odd quantities of the order
parameter
in general,
such systematic errors do not cast any serious problem,
because one can calculate $\chi_m$ as just the square of $m$ at
$T\geq T_c$
(there can be no unique definition of $\chi_m$ at $T < T_c$ for
finite systems).
On the other hand, for the response functions of even quantities,

such as the specific heat,
one always has to be careful about that systematic errors;
Even then, however, one can reduce the errors by making many
independent
runs and using Eq.~(\ref{eq:susest}).

\bigskip
\bigskip
\noindent{\bf Acknowledgements:}
Two of the authors (M.~K. and N.~I.) would

like to thank Professor K.~Binder and Professor D.~Stauffer
for their kind hospitality, fruitful discussions, and encouragement.
Thanks are also due to Dr. H.~Takano and Dr. T.~Koma for
valuable discussions.
The work was partially supported by a Grant-in-Aid

for Scientific Research on Priority Areas, ``Computational Physics

as a New Frontier in Condensed Matter Research'', from the Ministry

of Education, Science and Culture, Japan.


\begin{thebibliography}{99}
\bibitem{KI93} M.~Kikuchi and N.~Ito: J.~Phys.~Soc.~Jpn. {\bf 62}
(1993) 3052.
\bibitem{MB73} H.~M\"uller-Krumbhaar and K.~Binder: J.~Stat.~Phys.
{\bf 8}
(1973) 1.
\bibitem{IKO93} N.~Ito, M.~Kikuchi, and Y.~Okabe: Int.~J.~Mod.~Phys.
{\bf C4} (1993) 569.
\bibitem{I93} N.~Ito: Physica {\bf A192} (1993) 604 (also
ref.~\cite{KI93})
and references therein for 3D\\
N.~Ito: Physica {\bf A196} (1993) 591 and references therein for 2D.
\bibitem{WL91} S.~Wansleben and D.~P.~Landau: Phys.~Rev. {\bf B43}
(1991) 6006.
\bibitem{H92} H.~O.~Heuer: J.~Phys. {\bf A25} (1992) L567.
\bibitem{FLB91} A.~M.~Ferrenberg, D.~P.~Landau, and K.~Binder:

J.~Stat.~Phys. {\bf 63} (1991) 867.
\end{thebibliography}
\end{document}